\documentclass[sigconf]{acmart}
\usepackage{multirow}
\AtBeginDocument{%
  \providecommand\BibTeX{{%
    \normalfont B\kern-0.5em{\scshape i\kern-0.25em b}\kern-0.8em\TeX}}}

\copyrightyear{2023} 
\acmYear{2023} 
\setcopyright{rightsretained} 
\acmConference[AIES '23]{AAAI/ACM Conference on AI, Ethics, and Society}{August 8--10, 2023}{Montréal, QC, Canada}
\acmBooktitle{AAAI/ACM Conference on AI, Ethics, and Society (AIES '23), August 8--10, 2023, Montréal, QC, Canada}
\acmDOI{10.1145/3600211.3604685}
\acmISBN{979-8-4007-0231-0/23/08}

\usepackage[printonlyused,nolist,nohyperlinks]{acronym}
\usepackage{graphicx}
\usepackage[title]{appendix}
\usepackage{longtable}
\begin{document}
\acrodef{ML}{machine learning}
\acrodef{FMEA}{Failure Mode and Effects Analysis}
\acrodef{STPA}{System Theoretic Process Analysis}
\acrodef{T2I}{Text-to-image}
\acrodef{UCA}{Unsafe Control Action}
\newcommand{\am}[1]{\textcolor{red}{AJ: {#1}}}
%
\title{Beyond the ML Model: Applying Safety Engineering Frameworks to Text-to-Image Development
}


\author{Shalaleh Rismani}
\affiliation{%
 \institution{McGill University}
 \city{Montreal}
 \state{QC}
 \country{Canada}
}

\author{Renee Shelby}
\affiliation{%
  \institution{Google Research}
  \city{San Francisco}
  \state{CA}
  \country{USA}
}

\author{Andrew Smart}
\affiliation{%
  \institution{Google Research}
  \city{San Francisco}
  \state{CA}
  \country{USA}
}

\author{Renelito Delos Santos}
\affiliation{%
  \institution{Google Research}
  \city{San Francisco}
  \state{CA}
  \country{USA}
}

\author{AJung Moon}
\authornote{Senior authorship is shared between the last two authors, AJung Moon and Negar Rostamzadeh.}

\affiliation{%
  \institution{McGill University}
  \city{Montreal}
  \state{QC}
  \country{Canada}
}

\author{Negar Rostamzadeh}
\authornotemark[1]
\affiliation{%
   \institution{Google Research}
   \city{Montreal}
    \state{QC}
   \country{Canada}
}
\begin{abstract}
Identifying potential social and ethical risks in emerging \ac{ML} models and their applications remains challenging. 
In this work, we applied two well-established safety engineering frameworks (FMEA, STPA) to a case study involving text-to-image models at three stages of the \ac{ML} product development pipeline: data processing, integration of a T2I model with other models, and use. Results of our analysis demonstrate the safety frameworks -- both of which are not designed explicitly examine social and ethical risks -- can uncover failure and hazards that pose social and ethical risks. We discovered a broad range of failures and hazards (i.e., functional, social, and ethical) by analyzing interactions (i.e., between different \ac{ML} models in the product,  between the \ac{ML} product and user, and  between development teams) and processes (i.e., preparation of training data  or workflows for using an ML service/product). Our findings underscore the value and importance of examining beyond an \ac{ML} model in examining social and ethical risks, especially when we have minimal information about an \ac{ML} model. 

\end{abstract}

\ccsdesc[500]{General and reference~Evaluation}
\ccsdesc[500]{Computing methodologies~Machine learning}

\keywords{Safety engineering,
T2I generative models,
Responsible ML, Art}



\maketitle

\section{Introduction}
\label{introduction}

Scholarly work reveals ML-based products and services can facilitate and scale discriminatory treatment of marginalized groups \cite{Tomasev2022-cg}, spread misinformation \cite{Weidinger2022-qv}, and deteriorate a user's sense of autonomy \cite{Carli2022-yv}. Such negative outcomes present themselves as \textit{social and ethical risks} to direct and indirect stakeholders of these technologies. Identifying, assessing, and mitigating such risks, however, is challenging for practitioners. To intervene in these concerns, researchers have proposed quantitative \cite{Franklin2022-qe, Lee2021-oo}, qualitative \cite{Liao2020-cy, Gebru2021-kx, Mitchell2019-ms, Raji2020-dw}, and epistemological frameworks \cite{Ehsan2021-st, Mohamed2020-uu, Hanna2020-rn} to better identify and manage the social and ethical risks of \ac{ML} systems; however, many proposed evaluation methods focus narrowly on the performance and properties of a single ML model (i.e. fairness metrics) as opposed to examining the associated processes and systems. Recently, there is increased attention paid to social and ethical risks that arise in ML development processes (i.e., data collection practices \cite{Sambasivan2021-nv}) and interactions (i.e., contextual use of an ML system \cite{Wang2022-ck}). However, empirical studies with responsible ML practitioners find existing approaches to assessing social and ethical risks of a single ML model or relevant processes are often implemented at ad-hoc basis, \cite{Rismani2022-ao, Madaio2022-yz,Rakova2021-ov}. Furthermore, besides organizational-level challenges of inadequate incentives and resources \cite{Widder2022-rf,Mantymaki2022-im}, many practitioners tasked with managing the social and ethical risks of an \ac{ML}-based product or service have minimal understanding of the underlying ML model(s) due to their technical complexity and the often-inadequate documentation/communication practices between variety of people or teams involved in ML model development from data collection to productionization \cite{Rismani2022-ao, Moss2020-le}. Considering these challenges, practitioners have emphasized the need to establish systematic and structured approaches for social and ethical risk management of ML-based products \cite{Rismani2022-ao}.

Given its focus on structured risk-reduction, scholars in the \ac{ML} community have argued for the use of safety engineering frameworks -- particularly \ac{STPA} and \ac{FMEA} -- as means to analyze and manage social and ethical risks of \ac{ML} systems \cite{Dobbe2022-kf,Raji2020closing,Rismani2021-qy, Khlaaf2022-rz}.
These two frameworks, in particular, are well-established in the safety engineering practice and have been used in the design and development of safety-critical systems since the 1940s \cite{leveson2016engineering, Carlson2012-yh}. Recent scholarly work highlights that these frameworks could provide the necessary systematic structure for assessing the social and ethical risks of ML systems \cite{Rismani2022-ao,Khlaaf2022-rz,Rismani2021-qy,Dobbe2022-kf}. However, there remain open and unexplored questions on how we can apply safety frameworks (e.g., what aspects of the ML development pipeline should be considered in the analysis scope) and what these frameworks can reveal about potential social and ethical risks of these ML applications. 

As social and ethical risks often emerge from both how a technology is developed and how it is embedded within a social context \cite{sambasivan2021re, Wang2022-ck, Friedman2019-jp}, examining contextual aspects provides valuable insight for risk management and harm reduction. Safety engineering frameworks are frequently used to examine harms that could emerge from a process (i.e., a manufacturing process) or interactions between different sub-parts of a system (i.e., between different internal components) \cite{Carlson2012-yh,Leveson2018-no}. These frameworks have the necessary analytical approach for connecting potential failures in a process or an interaction to downstream harms. We leverage the capability of these frameworks to examine interactions and processes involved in the ML-product/service development pipeline and investigate if they enable the discovery of social and ethical risks without changing the original frameworks. Our research questions are: 
\begin{itemize}
    \item RQ1: What can \ac{STPA} and \ac{FMEA} reveal about social and ethical risks by examining processes and interactions involved in developing and deploying an ML model? 
    \item RQ2: How can \ac{STPA} and \ac{FMEA} be conducted along the \ac{ML} development pipeline to identify potential social and ethical failures/hazards of a system?

\end{itemize}
We focus on the application of \ac{FMEA} and \ac{STPA} at three stages of the \ac{ML} development and deployment pipeline: (1) data processing for creating a training dataset, (2) integration of one ML model with other \ac{ML} or non-\ac{ML} algorithms in an ML-product, and (3) end use of the ML product. To illustrate, we conducted FMEA and STPA on a case study involving real-world users of a text-to-image (T2I) model user interface by professional visual artists in their creative practice. The rapid public release and adoption of large generative models in various application areas have fueled much concern about the societal and ethical implications of generative models (e.g., \cite{disinfo2022, Struppek2022-au, Qiao2022-fu}). Considering the complexity of these models, investigating ML development and deployment process could provide a vantage point for identifying social and ethical risks \cite{Gero2022-dv}.  
This analysis is solely illustrative and it captures a point-in-time snapshot and potential configuration of the development and deployment for the chosen ML application. While we did not conduct a full STPA/FMEA analysis, this case study offers empirical evidence on how safety engineering frameworks could be translated to analyzing social and ethical risks. 

Our analysis illustrates that even without having detailed information about the \ac{ML} model, safety engineering analysis provides a systematic method of discovering a range of failures and hazards along the ML development and deployment pipeline. We discovered potential failures and hazards that pose social and ethical risks by analyzing \textit{processes} and \textit{interactions} surrounding a given generative model in an ML product, even though these frameworks were not originally designed for uncovering such risks. STPA and FMEA provided a systematic and consistent approach to analyzing a range of interactions and processes including 1) process of training an ML model, 2) interaction between an ML model and accompanying models in a given product, 3) interaction between an ML product and its users. Lessons learned from our analysis can guide practitioners in conducting systematic analysis beyond a single model, which reflects the majority of production use cases at organizations. With the rapid adoption of \ac{ML} models in various products today, we call for further examination and use of safety engineering frameworks to improve responsible \ac{ML} development and integration despite the increased opacity and complexity of the models involved.

In the remainder of this paper, we contextualize the relevance of the safety engineering frameworks and provide an overview of current discourse in T2I model development and use in the creative process (Section \ref{background}). We outline our information-gathering protocol and analysis methods (Section \ref{methodology}) and then highlight key findings (Section \ref{findings}). Lastly, we discuss the value and shortcomings of applying safety engineering frameworks in light of current practices and call on the research community to further examine and strengthen these frameworks for ethical and social risk management of \ac{ML} systems (Section \ref{discussion}).

\section{Background}
\label{background}
Practitioners and scholars in the responsible ML field have proposed a range of tools and frameworks for identifying, assessing, and mitigating potential social and ethical risks of ML systems. These tools include approaches for assessing the properties of models with respect to values such as fairness or transparency \cite{Madaio2020}, examining the interaction between a model and its context of use \cite{Raji2020-dw}, and investigating the ML model creation processes \cite{sambasivan2021re}. Many of the current assessment methods are applied in an ad-hoc basis across the ML development pipeline limiting their impact and in response, practitioners have expressed the potential for safety engineering to provide a systematic approach for managing social and ethical risks of ML systems \cite{Rismani2022-ao, Dobbe2022-kf}. In the following sections, we provide a brief description of scholarly work that discusses the potential use of safety engineering frameworks for ML systems and discuss existing studies relevant to our case study on the use of T2I models in artistic creations. 

\subsection{Safety engineering frameworks: failure and hazard analysis}
Safety engineering is a long-standing discipline that has evolved from creating safe mechanical systems (ex., planes, and cars) to safe software systems \cite{Dekker2019-sk}. As \ac{ML} systems can scale both benefits and harms, there is a need to investigate how existing safety engineering frameworks could support safe \ac{ML} development and deployment \cite{Khlaaf2022-rz,Rismani2022-ao,Raji2020-dw,Khlaaf2023-qs}. Failure and hazard analyses are the most commonly used frameworks in safety engineering practice \cite{Dekker2019-sk}. This type of analysis is often conducted early in the development process to foresee potential failures and develop ways to control them in design \cite{Carli2022-yv,leveson2016engineering}. These methods differ from other assessment processes for \ac{ML} systems, such as algorithmic impact assessments and auditing practices, in their focus on identifying, evaluating, and connecting anticipated harms to a design decision and mitigation in the development process \cite{Raji2020-dw,Rismani2022-ao}.

Prior work in responsible ML development has discussed the potential use of two failure and hazard analyses processes, \ac{FMEA} and \ac{STPA}, for managing social and ethical risks \ac{ML} systems ~\cite{Li2022-vt,Dobbe2022-kf,Rismani2021-qy,Raji2020-dw}. FMEA is a well-established reliability engineering process \cite{Jenab2015-ts,Hindawi_undated-gp} that takes an analytical reduction approach (i.e. breakdown of a system into its steps or components) to evaluate the likelihood of risk for potential failure modes \cite{Carlson2012-yh}. On the contrary, STPA is a hazard analysis tool that uses a system theoretic perspective to map out parts of a system and how they interact with each other \cite{leveson2016engineering}. Through this examination, analysts can identify potential hazards (i.e., sources of harm) and develop necessary safety requirements \cite{Ishimatsu2014-ni,Shin2021-tg,ishimatsu2010modeling}. In contrast to FMEA, the STPA process does not focus on the likelihood of risk or specific points of failure.  Instead, STPA models and examines elements of control and feedback in a sociotechnical system. 

Existing research outlines the overarching benefits of FMEA for internal \ac{ML} auditing ~\cite{Raji2020-dw}, interprets how FMEA could reveal \ac{ML} fairness-related failures ~\cite{Li2022-vt}, and employs FMEA to suggest an analysis of "social failure modes" for \ac{ML} systems ~\cite{Rismani2021-qy}. Similarly, several works discuss the benefits of a system theoretic perspective for addressing social and ethical risks of \ac{ML} systems by allowing the analysts to map out how an ML system interacts with its environment ~\cite{Dobbe2022-kf,Martin2020-dl}. Recent studies explore industry \ac{ML} practitioners’ perspectives towards safety engineering techniques and highlight that safety engineering frameworks provide an avenue for systematizing the identification and mitigation of social and ethical risks ~\cite{Martelaro2022, Rismani2022-ao}. However, both studies recognize that a successful translation of these frameworks for ML development requires organizational changes/support and further empirical examination/development of these methods. In this work, we focus on the second gap and investigate how these two safety engineering frameworks could be used to identify social and ethical risks along the ML development process via a case study application. We focus our case study on the use of T2I models in the art creation process.

\subsection{Use of T2I models in creative practice}
In recent years, scholars and practitioners have developed highly performative models that generate images from a given text prompt. Such models, including DALL-E \cite{radford2021}, Parti \cite{Yu2022-ed}, Stable Diffusion \cite{stablediff}, and Imagen \cite{Saharia2022-em} are generative T2I models. They perform significantly better in terms of image quality and text-to-image alignment compared to their predecessors \cite{Yu2022-ed,radford2021}. These T2I models have been released for general public use through various user interfaces and APIs \cite{Davenport2022-vt, Zwiezen2022-lv, Johannezz2021-jm}.

Artists, especially digital artists, have been among the early adopters of many T2I user interfaces \cite{promptbattle} leading to a growing discussion on how artists could use T2I models for co-creation ~\cite{Guzdial2019-pc}. Alongside enthusiasm in certain artist communities, however, there is growing concern about the potential harm that could emerge from their use. This includes artist concerns about how their artwork is often used as training data in the creation of such models \cite{edwards2022}, how generative models could affect art creation practice \cite{clarke2022}, and impact artists’ livelihood \cite{salkowitz2022, Chng2019-xs, Hertzmann2020-nj}. Similarly, ML and Responsible AI scholars have examined how image generation models could perpetuate existing systematic biases \cite{Wang2022-ck, Struppek2022-au, Bandara2022-sl, Tomasev2022-cg, Birhane2021-kh, Cho2022-mw}, including stereotype amplification. Creators of such models have recognized potential limitations and risks posed by these models in their public releases of academic papers and APIs \cite{Yu2022-ed}.

\subsubsection{Rationale for choice of case study} 

We focus our case study on the use of T2I models in the context of art creation. Despite significant improvements in specific performance metrics (i.e., image quality and text-to-image alignment) in recent years \cite{Yu2022-ed}, large generative models, such as T2I models, are opaque and complex, making it challenging to uncover potential failures and hazards \cite{Bommasani2021-wm}. However, practitioners still consider their use across many applications and use cases \cite{Arora2023-em,Guzdial2019-pc}. Safety engineering analysis allows practitioners to look beyond the properties of a single complex model and discover potential failures/hazards by investigating the processes that are part of the development and deployment of such models. We posit that this approach is especially beneficial for assessing the risks of more generalized models and empirically examine this by choosing a case study around the use of T2I models. Even though safety analysis is often conducted for safety-critical systems (i.e., nuclear power plants, airplanes, medical devices) \cite{Dekker2019-sk}. By choosing a case study on the use of T2I models in creative practice, we leverage these systematic approaches for risk assessment and harm reduction in applications that have emerging hazards/failures \cite{Wang2022-ck, Cho2022-mw} but are not categorized formally as safety-critical. 

\section{Methodology}
\label{methodology}
We use a case study approach to explore our two research questions described in Section \ref{introduction} for the following reasons \cite{Rashid2019-iz}. First, there is a lack of precedence in how such tools could be applied to ML systems. Second, empirical evidence is needed to understand the nature of failure and hazards that emerge from such analysis to see if these tools could allow us to uncover potential social and ethical risks. While many variations of STPA/FMEA exist, the first step of traditional STPA and FMEA requires mapping certain information about a given system \cite{Carlson2012-yh, leveson2016engineering}. Therefore, we gather the necessary information prior to conducting STPA and FMEA. 

\subsection{Information gathering}
Typically, FMEA and STPA are conducted by system experts and safety engineers who are working in a company. These analysts have in-depth knowledge of the systems and the safety engineering processes. Considering that both the FMEA and STPA are conducted by the authors of this paper, we needed to gather information about the system, its components, and its interaction with various stakeholder groups. This is necessary in order to divide the system into functional components or steps (in the case of FMEA), and losses and constraints (in the case of STPA). 

In our case study, we collected three different sources of data for conducting FMEA and STPA analysis (as illustrated in the supplemental material) including workshops with artists, expert interviews with T2I model developers/evaluators, and secondary research on T2I models. We describe our process for gathering information from the workshops and the interviews in the following sections and follow by describing our analysis approach. 

\subsubsection{Workshop with artists}
To understand the artist’s perspectives towards the use of T2I models in their creative process, we conducted three 90 minutes-long workshops with 15 artists. 

\textit{Participant recruitment:} We used purposive \cite{palinkas2015purposeful} and snowball sampling \cite{parker2019snowball} to recruit participants for this workshop. The workshop organizers brainstormed an initial list of artists and only included candidates that had worked with T2I models in their practice, were older than 18-years-of-age and were professionally working in the arts for at least a year. Participants were recruited via email and once they accepted to participate, they were sent a consent form. In total 15 artists representing 6 countries participated in the workshop. Participants held a diverse set of roles in the creative industry, including but not limited to, filmmaker, art curator, and digital artist. 

\textit{Workshop protocol:} The workshop protocol (as illustrated in the supplemental materials) included three different sections, each of which was 90 minutes long. The first section focused on getting to know each participant and how they use T2I models within the creative process. In the second section, one of the researchers presented a sociotechnical harms taxonomy \cite{Shelby2022-oi} and facilitated a discussion on the perceived harms of T2I models with the group of artists. The third section of the workshop focused on discussing potential harm reduction avenues.

\subsubsection{Interviews with T2I developers and evaluators}
We conducted 60-min interviews with 8 industry experts involved in the development and evaluation of T2I models to understand their processes. 

\textit{Recruitment:} Similar to the workshop recruitment, we used purposive \cite{palinkas2015purposeful} and snowball sampling \cite{parker2019snowball} to recruit interview participants. We brainstormed an initial list of candidates who were 18 years or older and had worked on the development or evaluation of a T2I model for at least 1 year. The researchers reached out to potential candidates via email and a consent form was sent to individuals who agreed to participate in the interview. Overall 8 people participated in the interview covering three roles: 1) Researchers who worked on developing and evaluating T2I models for performance, social and ethical issues, 2) Software engineers who developed parts of the T2I demo, and 3) Managers who coordinated the release of T2I demos. 

\textit{Interview protocol:} The interview protocol, as illustrated in the Supplemental Material, was designed to understand the process of development and evaluation the participant used when working on T2I models and outlined the interaction between different stakeholders on a given T2I model project. All interviews were 60 minutes in length.

\subsection{Conducting FMEA and STPA}

FMEA and STPA could be applied to various scopes of analysis that require a different set of information. Scoping an \ac{ML} system for such an analysis is a non-trivial task. An \ac{ML} system can be divided into its component parts (e.g., training data, model, user interface), development process (e.g., training, testing, early deployment), stakeholders involved (e.g. ,\ac{ML} developers, a community of developers interfacing and building on the model APIs, end-users), and so on. Results of the analysis can be drastically different depending on the chosen scope of analysis \cite{Carlson2012-yh,leveson2016engineering}. 

Based on the information gathered from the interviews with T2I experts and workshops with artists, we identified three scopes of analysis along the \ac{ML} development pipeline: 
\begin{itemize}
    \item Scope 1: the data processing necessary for creating a training dataset 
    \item Scope 2:  how a T2I model is integrated into a production environment along with other \ac{ML} models
    \item Scope 3: how an artist uses T2I model demo as part of their creative process 
\end{itemize}

Selecting the scopes along the \ac{ML} development pipeline allows for the examination of critical processes and interactions as discussed by previous scholarship \cite{Hutchinson2020-jm,sambasivan2021re,Rismani2022-ao}. We chose to focus on these scopes of analysis because we had access to the most amount of publicly available and shareable information about the elements involved. We recognize the value of other scopes such as how a model or a product is evaluated, or how the model architecture is designed, and encourage that scopes beyond what we have experimented with in this paper are considered for future applications of FMEA and STPA.

The STPA and FMEA analysis was led by the first author of this paper and three of the co-authors provided feedback on iterations of the analysis. A separate analysis was conducted for each one of the scopes. In total three FMEA and three STPA analyses were conducted. The lead author spent somewhere between 10 - 12 hours implementing the FMEA or the STPA process on one scope. All sources of data were used as input for both the STPA and FMEA analysis. The FMEA process resulted in a list of potential failure modes, and a Risk Priority Number score. The STPA analysis resulted in a list of unsafe control actions and corresponding safety requirements. We followed the original STPA and FMEA process for each one of the scopes (as described below) and did not alter them to specifically uncover social and ethical risks.




\subsubsection{FMEA process}

FMEA is a multi-step framework, through which steps are iteratively performed by FMEA and system experts over the development life cycle ~\cite{Carlson2012-yh} (refer to Figure \ref{fig:fmea}): 

\begin{figure*}[ht]
  \centering
    \Description{The figure illustrates the most commonly performed steps for FMEA} 
\caption{Steps for conducting an FMEA \cite{Carlson2012-yh}}
  \label{fig:fmea}
  \includegraphics[width=0.8\linewidth]{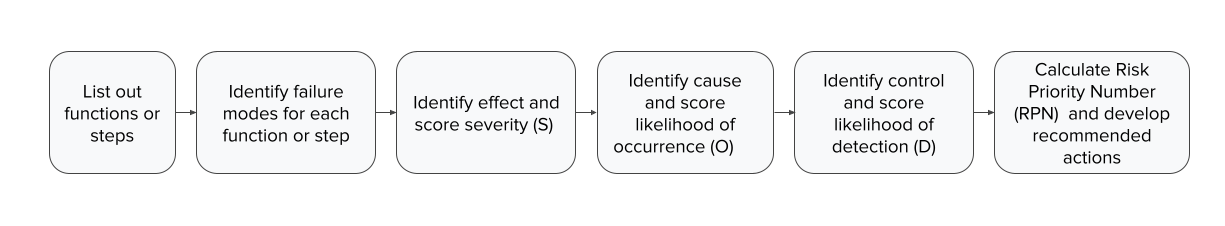}
\end{figure*}

\begin{enumerate}
    \item List out the \textit{functions} of a component/system OR steps of a process (e.g., everything the system/process needs to perform). 
    \item Identify potential\textit{ failure modes}, or mechanisms by which each function or step can go wrong.
    \item Identify the \textit{effect}, or impact of a failure, and score its \textit{severity} on a scale of 1 – 10 (least to most severe).  
    \item Identify the \textit{cause}, or why the failure mode occurs, and score its \textit{likelihood of occurrence} on a scale of 1 – 10 (least to most likely). 
    \item Identify \textit{controls}, or how a failure mode could be detected, and score \textit{likelihood of detection} on a scale of 1 – 10 (most likely to least likely). The scales used in the automotive industry standards \cite{aiag} (illustrated in the supplemental material) were used for scoring severity, the likelihood of occurrence, and the likelihood of detection. 
    \item Calculate \textit{Risk Priority Number} (RPN) by multiplying the three scores;  a higher RPN indicates a higher risk level and develops \textit{recommended actions} for each failure mode and prioritize based on RPN. 
\end{enumerate}

\subsubsection{STPA process}
\begin{figure*}[ht]
  \centering
    \caption{Steps for conducting an STPA \cite{Leveson2018-no}}
  \label{fig:stpa}
  \includegraphics[width=\linewidth]{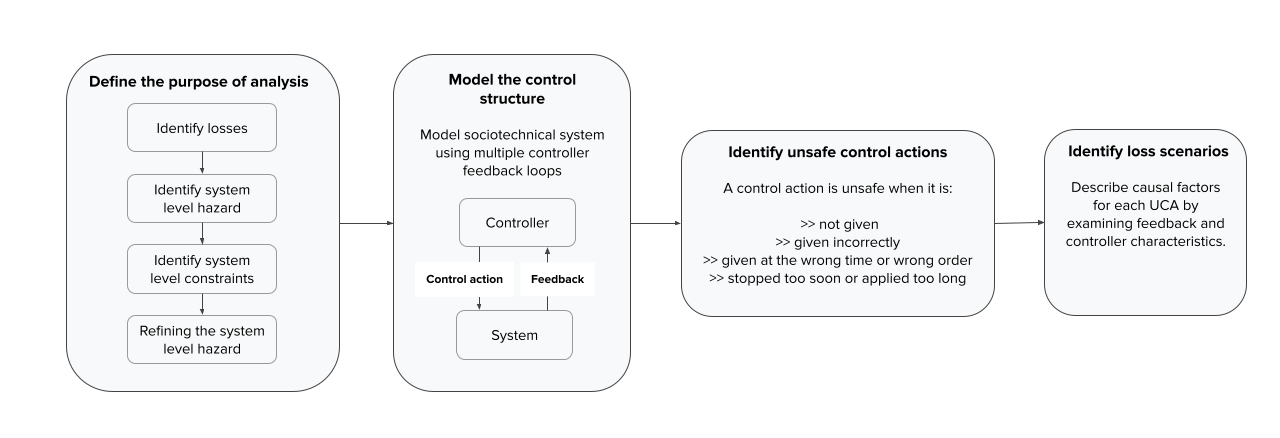}
\end{figure*}

STPA is a hazard analysis framework that is performed and led by system and safety experts, iteratively (across the model of a system) and cyclically (across a system’s lifecycle) (refer to Figure \ref{fig:stpa}).

\begin{enumerate}
    \item Define the \textit{purpose of the analysis} by identifying losses via outlining stakeholders and their values. System-specific hazards and controls are highlighted based on the specified loss.  
    \item Model the \textit{control structure} of the full sociotechnical system using control feedback loops which consists of a controller which sends \textit{control actions} to a system that is being controlled while receiving \textit{feedback} from the same system. 
    \item Identify \textit{unsafe control actions} (UCA) by going through each control action and thinking about unsafe modes of (no) action, incorrect action, and untimely action. 
    \item Identify potential \textit{loss scenarios} by outlining potential causal scenarios (i.e., missing feedback loops, incorrect process model or control algorithm of a controller) for each UCA.
\end{enumerate}

\subsection{Method of analysis}
After conducting FMEA and STPA analyses for the three different scopes we examined two key elements for each scope: (1) Could any of the identified failures and hazards lead to social and ethical risks? (2) Did our analysis discover any new issues that have not been reported in the literature or media? We then reflected on common themes and lessons learned that appeared across all six applications. 

\subsection{Author reflexivity and limitations}
All the authors of this paper are living and working in institutions in the Global North. We recognize our lived experiences and perspectives impact our choice to use safety engineering frameworks (i.e. safety is valued and practice differently across the world), methodology, and outcome of this analysis. Out of our authorship,  four individuals have conducted FMEA and STPA in training programs or as practitioners in the industry. However, none of the authors are experts with 10+ years of experience in safety engineering. Furthermore, the STPA and FMEA we completed are based on reported information gathered in the interviews and workshops. They do not capture the direct opinion and knowledge of the stakeholders, as these experts are not participating fully in the workshops. Our proof-of-concept analysis could be improved with the presence of expert safety engineering practitioners and system (i.e., T2I) developers/evaluators. Finally, this paper focuses on one case study, which limits the generalizability of the findings across ML models. 

\section{Key Findings}
\label{findings}

Applying FMEA and STPA at three stages of the \ac{ML} development pipelines enabled analysts to uncover a wide range of failures and hazards without the need for detailed information on the specific T2I model. In particular,  the three scopes of analysis (training data processing, product integration, and end use) each allowed practitioners to look beyond a specific model in isolation and focus on \textit{processes} and \textit{interactions} as it is integrated with other actors and technical systems. In the next three sections, we reflect on the failures and hazards discovered at each stage. The analysis uncovered known failures and hazards (e.g., creation of non-consensual sexual imagery) and novel ones that have not been recognized to the best of our knowledge (e.g., the impact of English-word filters on lexical change). We also identified failures and hazards that could present social and ethical risks for different actors, which can inform prioritization of mitigation strategies. The analyses presented in this paper are proof-of-concept and illustrative examples of how safety frameworks could be applied at different stages of the ML development pipeline. They are not meant to be a comprehensive failure and hazard analysis for the use of T2I models by artists. The identified failures and hazards would shift depending on the assumptions used to set up the analysis. Therefore, it is important to interpret our findings as an illustration of how FMEA and STPA could be applied by practitioners, rather than a final analysis. In what follows, we discuss the process of applying STPA and FMEA process at different scopes, the nature of our findings, highlighting a sample of the identified failures and hazards in Tables \ref{tab:fmea-data} - \ref{tab:stpa-use}.

\subsection{Scope one: data processing for creating training dataset}
\label{scope1}
In our application, we bound the data processing stage to start from the identification of one or multiple source datasets and end with the creation of a training dataset ready for model development purposes. In this scope, we do not directly analyze the steps involved in how the data sources were obtained and who was involved in the data collection process. The focus is on how the data sources are processed for creating a training dataset. From our interviews, we identified three key stakeholders involved in data processing: (1) software engineers who prepare the dataset for training; (2) responsible \ac{ML} practitioners who provide guidelines on what should or should not be included in the dataset; and (3) lawyers who consult on privacy and legal requirements for datasets. The software engineers collaborate closely with the practitioners responsible for developing the model. Lawyers provide legal requirements (i.e., intellectual property and privacy) for training datasets to the responsible ML and the product team. There is limited information available on how data processing is done for T2I models. For our proof-of-concept analysis, we use the data processing steps outlined in the publicly available data card for the Parti model \cite{fit400M}. The steps outlined in this data card are similar to what other T2I model creators have discussed in their publications \cite{Saharia2022-em, dalle}. 

\paragraph{\textbf{FMEA}} 
The FMEA analysis, as illustrated in Table \ref{tab:fmea-data}, starts with outlining the data processing steps \cite{fit400M}, which include: (1) filter for records identified as containing sensitive data; (2) filter for non-English data; (3) filter for "adult" content in images; (4) filter for text associated with "adult" content; (5) exclude low text-image semantic alignment; and (6) exclude text consisting of mostly numbers. We identified 3 to 5 failure modes for each step, resulting in 22 failure modes for this scope of analysis. We identified six failure modes due to under-performance (i.e., some sensitive data is not captured), six failures modes due to over-performance (i.e., English data is filtered alongside non-English data), or six failure modes due to loss of performance (i.e., low text-image semantic alignment is included) of the filtering function. Four of the failure modes were due to unintended behavior of the filtering functions, such as when a non-English language filter does not recognize more modern English words. When taking a closer look at the nature of these failure modes, ten describe performance-related issues around image quality and text-to-image alignment. For example, over-filtering the source datasets is a known failure mode and could result in a lack of training data for creating a highly performative model. T2I developers have widely highlighted the importance of large training datasets in generating high-resolution images \cite{Yu2022-ed, dalle,Saharia2022-em}. 

Twelve of the identified failure modes have clear social and ethical implications. For instance, under-filtering the source data or missing filters could lead to the downstream generation of sensitive data or adult content, which could lead to interpersonal harms (i.e., non-consensual sexual imagery and related mental health or reputational impacts) \cite{Shelby2022-oi}. Some of the identified failures have been recognized in the literature or reported in the media. For instance, Birhane et al. discuss that many of the existing adult content filters are not able to fully detect and eliminate the target content \cite{Birhane2021-ry}. 

Notably, our preliminary proof-of-concept, allowed us to pinpoint potential failure modes that are not discussed widely or publicly for applications of T2I. One example failure mode is that "the non-English word filter eliminates English words that are emerging/new or used in specific social groups." Elimination of novel words (neologisms) in training data influence what the T2I model can/cannot generate. As lexical change often does come from historically marginalized groups, this failure mode could ultimately alienate artists from specific cultural and social groups from using the T2I demo in their creative practice. By conducting this type of analysis on the process (and not the model itself), a practitioner can identify both known and novel failures (and their resulting harms)that can emerge from training data processing choices.  
\begin{table*}
\caption{Sample proof-of-concept FMEA, data processing}
\label{tab:fmea-data}

\resizebox{\textwidth}{!}{%

\begin{tabular}{p{.13\textwidth}p{.1\textwidth}p{.25\textwidth}p{.24\textwidth}p{.24\textwidth}p{0.24\textwidth}}
\toprule

\textbf{Function} &
  \textbf{Type of failure mode} &
  \textbf{Failure mode} &
  \textbf{Effect} &
  \textbf{Cause} &
  \textbf{Control} \\ \hline
Filter records identified as containing sensitive data &
  Loss &
  Sensitive data is not filtered and included in the training set. &
  Sensitive data might be reproduced by the model output. &
  No requirement to filter data for sensitive data. &
  Model development team checks the data was filtered for sensitive information prior to starting the training. \\ \cline{2-6}
\textbf{} &
  Partial &
  Sensitive data is inadequately filtered and included in the training set. &
  Sensitive data is not filtered and it is included in the training set. Sensitive data might be reproduced by the model output. &
  Wrong filter thresholds are set. &
  Model development team monitors the outputs of the models for sensitive data generation. \\ \cline{2-6}
\textbf{} &
  Exceeding &
  Non-sensitive data is also filtered. &
  Resulting training dataset is too small. &
  Wrong filter thresholds are set. &
  Model development team monitors how much data is filtered. \\ \bottomrule
\end{tabular}}
\end{table*}

\paragraph{\textbf{STPA}}
The STPA process starts by identifying losses. To identify the values and losses associated with this scope of analysis, we reflect on the information gathered from the interviews and workshops to understand stakeholders impacted by the data processing stage. Artists emphasized the values of fostering creativity, serving a diverse audience, accessibility of artistic mediums to a diverse group of artists, efficiency in their creative process, and preservation of the artist's reputation/identity. The developers and evaluators of T2I models emphasized the value of creating efficient systems that generate appropriate and quality images in response to a given text. They also emphasized the value of diversity (i.e. the importance of serving a diverse audience of users with their models) and the importance of protecting their team and their company's public reputation. From the key values brought up by artists, software engineers, and responsible ML practitioners, we formulated the following losses: (L1) Loss of creativity; (L2) Loss of diversity; (L3) Loss of accessibility; (L4) Loss of efficiency; (L5) Loss of quality; and (L6) Loss of reputation.

The next STPA step is to identify hazards by considering these losses in relation to the previously identified goal of training data processing. We developed three potential hazards: 

\begin{itemize}
    \item H1: System creates a training dataset that contains low-quality text-image pairs. (L1, L4, L5, L6)
    \item H2: System creates a training dataset that contains harmful content. (L3, L5, L6)
    \item H3: System creates a training dataset that is not diverse in representation. (L2, L3, L4, L6)  
\end{itemize}

\begin{figure}[ht]
  \centering
    \caption{Control structure diagram, data processing}
  \label{fig:control-data}
  \includegraphics[width=\linewidth]{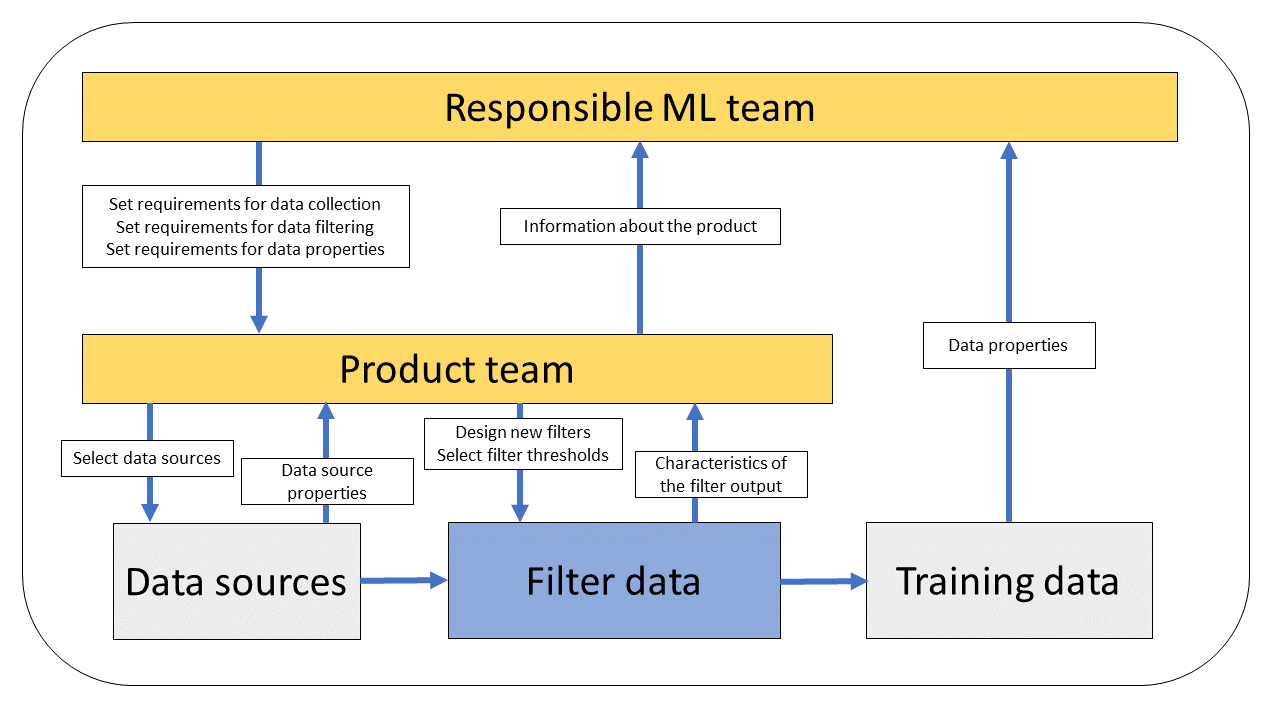}
\end{figure}

Then we model the control structure. A potential control structure configuration for the data processing stage, as illustrated in Figure \ref{fig:control-data}, includes two organizational/human controllers and one automated controller. We selected the "responsible \ac{ML} team" and "product team" as human controllers because they can make key decisions in the data processing scope. For this analysis, we assume the responsible \ac{ML} team is in charge of determining the parameters for a good training dataset and identifying key ethical and legal considerations. The product team is responsible for developing the models that underpin the T2I demo artists will use. We selected as "filter data" an automated controller to reflect where all of the automated filtering operations take place. We identified 6 control actions, denoted in the boxes with down arrows in Figure \ref{fig:control-data}.

For our analysis, we focused on 4 of the 6 control actions occurring between the three controllers and data sources including 1) Set requirements for data filtering, 2) Set requirements for dataset properties, 3) Design new filters and 4) Select filter thresholds. We chose to focus on these 4 control actions because they specifically focus on data processing as opposed to source data collection. We then brainstormed a list of 22 potential Unsafe Control Actions (UCAs) that relate to the three potential hazards we identified (partically illustrated in Table \ref{tab:stpa-data}). H1 focuses primarily on the functional performance of the system; whereas, the presence of H2 and H3 pose social and ethical risks. All 22 \acp{UCA} were linked to at least one hazard. This indicates the STPA process is attentive to the interconnected nature of emerging hazards, and can help practitioners identify how social and ethical risks are not separate from but directly related to performance hazards. 

Similar to the FMEA, we identified hazards related to improper filtering of data when examining the control action between the \textit{product team} and \textit{filter data} controllers, including over or under-filtration of the sensitive data. These \acp{UCA} could lead to all three hazard types (low-quality text-image pairs, harmful content, and homogenous representations). As discussed in the previous section, scholarly work has examined some of these \acp{UCA} in the context of T2I training dataset creation (e.g., under performance of adult content filters \cite{Birhane2021-ry}). While the STPA process revealed some of the same novel insights as the FMEA (i.e. incorrectly filtering English words created over time), the STPA results include additional \acp{UCA} related to how the human controllers (i.e. responsible ML team and the product team) interact with each other. For instance, one identified \acp{UCA} is that "the responsible \ac{ML} team provides the filtering requirements too late to the product team," preventing the product team from integrating the necessary filters. This is also one \ac{UCA} presenting social and ethical risks that is also linked to all three of the hazards. These types of \acp{UCA} may not be widely acknowledged in current literature because they focus on examining internal company processes (i.e., delayed communication internally), which can be overlooked in analyses focused solely on the model. Moreover, issues with internal company processes and practices are generally considered confidential information and hence they are not shared in the literature.  By identifying and addressing such \acp{UCA}, practitioners can embed responsible AI and safety considerations at an organizational level (i.e., beyond a single model or dataset). 


\begin{table*}
\caption{Sample proof-of-concept STPA, data processing}
\label{tab:stpa-data}
\resizebox{\textwidth}{!}{

\begin{tabular}{llllllll}
\toprule
\textbf{Controller} &
  \textbf{Controlled process} &
  \textbf{Control action} &
  \textbf{Type} &
  \textbf{Context} &
  \textbf{H1} &
  \textbf{H2} &
  \textbf{H3} \\ \hline
Responsible   ML   team &
  Product   team &
  Set   requirement for data filtering &
  Too   late &
  Responsible ML team provides requirements too late in the process &
  True &
  True &
  True \\ \hline
Product   team &
  Filter   data &
  Design   new filter for non-English data &
  Providing &
  The designed filter eliminate a large portion of data resulting in a small dataset &
  True &
  False &
  False \\ 
  \bottomrule
\end{tabular}}
\end{table*}

Both the FMEA and STPA frameworks allow practitioners to get a list of potential and plausible ways in which current data processing practices could fail and lead to potential social and ethical risks. The \acp{UCA} and the failure modes focused on shortcomings with filter design. However, STPA also revealed shortcomings in how requirements are communicated between groups. By conducting such analysis, teams could  keep track of how their data processing practices could fail and develop safeguards to ensure that training data is safely created for production-ready models.

\subsection{Scope two: Model/product integration}
Many AI ethics assessments and audits focus on one \ac{ML} model \cite{Ge2021-kb,Tahaei2023-hp}. However, in production, multiple models often are used to achieve the intended functions for a given product or demo \cite{Xin2021-ny}. We conducted FMEA and STPA to examine the integration of a T2I model in a productionized demo, such as those released by Stability AI \cite{stabilityai} and OpenAI \cite{dalle}. From our interviews with T2I developers and existing literature \cite{Rando2022-kb, Ngo2021-jn}, we identified demos often include at least three types of models: (1) an input prompt classifier (which either block or filter the text prompt), (2) the T2I model, and (3) output image classifiers (which either block or filter the generated image). For any given T2I demo, there could be multiple classifiers employed for filtering text prompts and images. For the purposes of this illustrative analysis, we assume there are only two classifiers. One for "adult" content run on the input text prompt and one for analyzing the generated image.

\paragraph{\textbf{FMEA}} We conducted a proof-of-concept system FMEA where we treat each of the three models as a sub-system, and perform the FMEA by first identifying the key function for each sub-system (see Table \ref{tab:fmea-model}). Considering our assumptions, we have three sub-systems in our product, with the following primary functions: (F1) Filter input text prompt for adult content; (F2) Generate 16 images per prompt; (F3) Filter generated images for "adult" content. 

We identified a non-exhaustive list of 14 potential failure modes for the three functions (3-5 failure modes for each function). These failure modes cover a range of issues, including functional failures, such as "no adult text prompt is filtered" or "it takes a long time to generate an image." The failure to generate images rapidly (i.e. latency) does not have obvious direct social or ethical risks; however, it affects company's reputation. 

\begin{table*}[t]
\caption{Sample proof-of-concept FMEA, model integration}
\label{tab:fmea-model}
\centering
\setlength{\tabcolsep}{4.5pt}
\resizebox{\textwidth}{!}{%

\begin{tabular}{p{.1\textwidth}p{.13\textwidth}p{.25\textwidth}p{.24\textwidth}p{.24\textwidth}p{0.24\textwidth}}

\toprule
\textbf{Function} &
  \textbf{Type of failure mode} &
  \textbf{Failure mode} &
  \textbf{Effect} &
  \textbf{Cause} &
  \textbf{Control} \\ \midrule
Filter   text prompts &
  Degradation &
  Filter does not work for novel words &
  Model does not filter prompts containing inappropriate language. &
  Safety filters are not updated/improved over time. &
  Product team monitors user feedback to identify potential negative feedback on prompt filters and complaints about the model accepting new and inappropriate prompts.  \\ \cline{2-6}
 &
  Partial &
  Filter works for English words written in Latin letters (e.g., it does not work when the Arabic language is written using Latin words) &
  Users enter harmful prompts using non-English phrases written with Latin letters &
  Filter allows text prompts containing non-English words &
  Product team monitors user feedback to identify potential negative feedback on how their prompts were filtered and complaints about the model accepting new and inappropriate prompts. \\ \cline{2-6}
 &
  Exceeding &
  Seemingly   appropriate prompts are rejected for no good reason. &
  Users   cannot enter what they would like to enter &
  Filters automatically eliminate word combinations that are not directly harmful but are in some form correlated to content that is marked as   harmful. &
  Product team monitors user feedback \\ \bottomrule
\end{tabular}}
\end{table*}

Similar to Scope 1, some identified failures for model/product integration present social and ethical risks. For instance, a complete loss of the filter function for the input or the output filters could harm potential users and pose social and ethical risks for the company and the artist, such as generating demeaning stereotypes. Many of the functional failures we identified had social and ethical implications as well, especially when considering the possibility that a given function might work well for some groups and poorly for others (i.e., quality-of-service harm). For example, partial filtering of the input prompts, and output images based on social norms of \textit{group A} exclusively could lead to differential performance for those from \textit{group B}. Similar risks emerge when the T2I demos only generate high-quality images for text prompts that represent terms, concepts, and ideas from a predominant social group (i.e., Western or Eurocentric cultures). 

Many failures we discovered in our proof-of-concept analysis (with its limited focus) have been discussed in recent literature examining potential representational and quality-of-service harms from T2I models \cite{Wang2022-ck, Struppek2022-au, Bandara2022-sl, Tomasev2022-cg, Johnson2022-hf}. However, in our analysis, we strictly followed the FMEA process and did not rely on the literature to identify these potential failure modes. Notably, our analysis shows that practitioners can identify potential failure modes by systematically following the FMEA process and considering the specific constraints/features of their own ML-based products to identify potential failure modes as opposed to solely relying on what has already been discovered in the literature.

\paragraph{\textbf{STPA}}
We start with the same set of losses outlined for the data processing scope as our key stakeholders for model integration. However, the hazards differ for this scope of analysis as the goal of the system has changed. Here the goal of the system is to create a T2I demo for public and creative use. Three potential hazards we considered are: 
\begin{itemize}
    \item H1: System creates an image that does not match the prompt. (L1, L3, L4, L5, L6)
    \item H2: System cannot generate an image. (L1, L2, L3, L4, L6)
    \item H3: System generates an unsafe image (i.e., with adult content). (L2, L3, L5, L6)
\end{itemize}

\begin{figure}[ht]
  \centering
  \caption{Control structure diagram, model integration}
  \label{fig:control-model}
  \includegraphics[width=\linewidth]{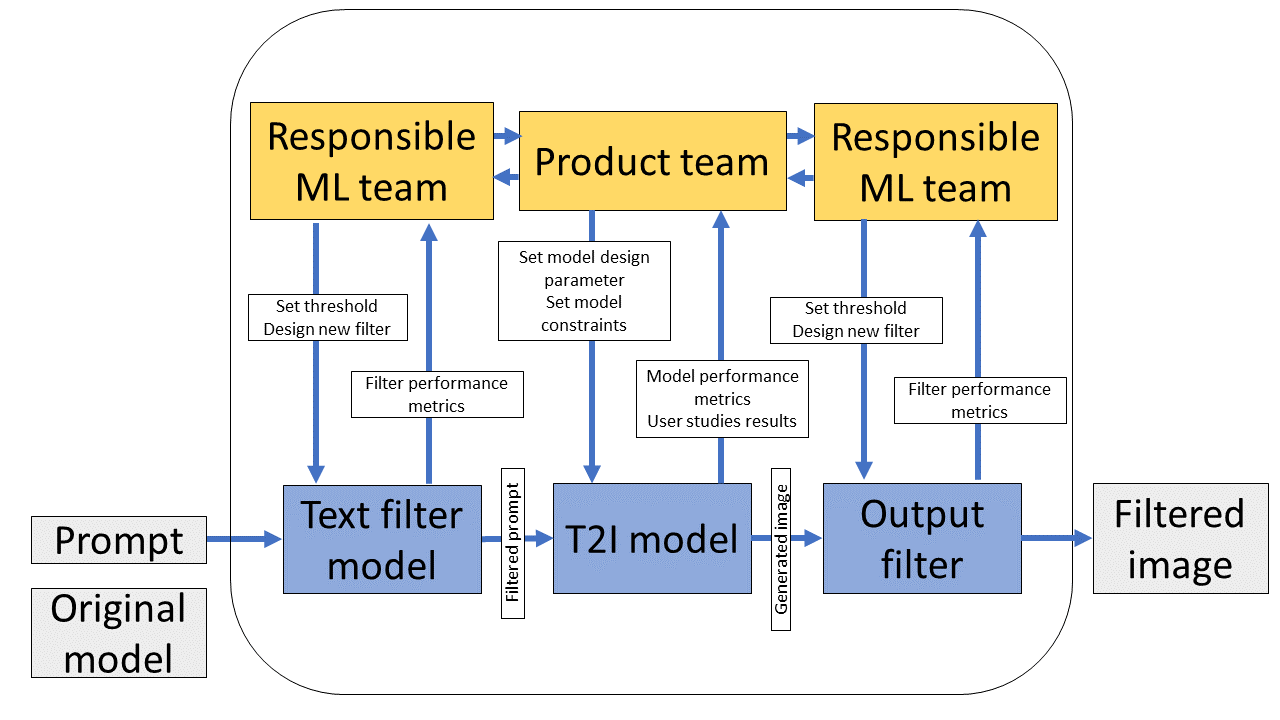}
\end{figure}

 A potential configuration of the control structure (as illustrated in Figure \ref{fig:control-model}) for this scope of analysis includes two organizational/ human controllers and three automated controllers. The organizational/human controllers are the "responsible \ac{ML} team" and the "product team." The automated controllers are the input text classifier, the T2I model, and the output image classifier. We identified 6 control actions, denoted in boxes with down arrows in Figure \ref{fig:control-model}. 

 For the purpose of this analysis, we focus on the 6 control actions between the two human controllers and the three models, which are: (1) Set a threshold for pre-designed filters for text and image filtering (counted as two control actions); (2) Design new filters for text and image filtering (counted as two control actions); (3) Set model design parameters; and (4) Set model constraints for the version of the T2I model used in this product (i.e. the T2I demo). We developed a non-exhaustive list of 16 \acp{UCA}. Table \ref{tab:stpa-model} highlights 2 potential \acp{UCA}. The \acp{UCA} capture process issues, such as delays in the delivery of requirements or miscommunication between the product and development teams. They also capture technical issues, such as incorrect filter thresholds or T2I model parameters. The 16 identified \acp{UCA} were often linked to more than one hazard. For example, the \ac{UCA} of "inappropriate filter threshold for the input text filter" could lead to H2 and H3. 
 
 All of the identified hazards could present a range of risks for stakeholders, including those posing social and ethical risk. For example, a system not being able to create an image for a subset of racially marginalized groups could present an \textit{ethical} and \textit{financial} risk for both the company and the artists trying to use the demo for their creative practice. As noted earlier, some of the identified \acp{UCA} are not well-discussed in the literature as they may emerge from internal organizational structures and configurations in their respective ML development pipelines. For example, a potential cause of an "incorrect model constraint" is a missing feedback loop between the output image filter and the product team. Based on the existing control structure, the product team only works on fine-tuning the T2I model for the demo. They do not see the results from the classifiers designed by the responsible \ac{ML} team. As illustrated, by examining the interactions between different models in an ML product and understanding how and who designs them, practitioners can identify and understand emerging hazards that would be ignored when focusing on one single model. 

\begin{table*}
\caption{Sample proof-of-concept STPA, model integration}
\label{tab:stpa-model}
\LARGE
\resizebox{\textwidth}{!}{

\begin{tabular}{llllllll}

\toprule
\textbf{Controller} &
  \textbf{Controlled process} &
  \textbf{Control action} &
  \textbf{Type} &
  \textbf{Context} &
  \textbf{H1} &
  \textbf{H2} &
  \textbf{H3} \\ \hline
Responsible   ML team &
  Input   filter &
  Design   new filters for the prompt filter &
  Too   late &
  New filters are not designed in time for product deployment &
  False &
  False &
  True \\ \hline
Responsible   ML team &
  Input   filter &
  Design   new filters for the prompt filter &
  Providing &
  New filters block most of the prompts that users would want  &
  False &
  True &
  False \\ \bottomrule
\end{tabular}}
\end{table*}


\subsection{Scope three: Use of the ML-based product}

This scope of analysis focuses on how an artist would use a T2I demo as part of their creative practice. The information we gathered from the artist workshops heavily informs this scope. To perform this illustrative FMEA/STPA analyses, we assume the artist is a filmmaker and they are creating a video for a client. The key stakeholders include the client, the filmmaker, and the public/viewer of the video. The scope focuses on how the filmmaker uses the T2I demo as a tool for generating images in their storyboard mock-up, where they visualize and share ideas with the client.  

\paragraph{\textbf{\textit{FMEA}}}
To investigate the potential failures that could come from the use of T2I demos by artists, we first mapped a potential \textit{process of use} based on the data collected in the workshops. All of the participants expressed that they use T2I demos as an image-generation tool within their artistic process to visualize ideas, communicate with collaborators and facilitate creative thinking.  Generalized from the artists' descriptions in the workshop, a  creative process for our assumed filmmaking scenario could involve the following steps: 
\begin{enumerate}
    \item Brainstorm storyboard ideas for advertising the product
    \item Develop prompts that represent the storyboard ideas 
    \item Enter prompts into the T2I demo to generate images 
    \item T2I demo generates image(s) based on given prompt 
    \item Select images for the storyboards 
    \item Share storyboards with clients/collaborators
    \item Integrate feedback and iterate to get a desired storyboard 
\end{enumerate}

Treating this workflow as our process of use, we conducted a proof-of-concept FMEA. To simplify the FMEA application, we narrowed the scope of analysis to steps 3 through 5, and identified a non-exhaustive list of 11 potential failure modes. The 11 failure modes encompassed potential ways in which artists cannot use the T2I demo in their workflow. For example, "the artist cannot enter prompts in their native language," "they can only use a limited set of words in the input prompt," or the "generated images did not match their expectations/needs." We also discovered technical failures of the T2I demo, such as "the generation of low-quality images" or "the generation of unsafe images (as identified in generated image placeholders)." Many of the failure modes we identified uncovered challenges of an artist with using T2I demo in the generation practice (i.e., low-quality image generation, not being able to generate an image or enter a prompt). These types of failure modes could mainly lead to performance-related risks. Similar to Scope 2 and 3, these performance-related risks could present social and ethical risks for some user groups who exclusively experience the effect of the failure (i.e., an artist cannot generate images related to the cultural concepts). Moreover, we identified a few failures that could present social and ethical risks directly, including when the T2I demo generates a harmful image or when an artist selects images that could be harmful to a specific audience. This could be a failure depending on the intent of the artist, as some art is meant to be politically provocative and hence has the potential to harm. The failure modes on performance issues of T2I demos and their ability to generate harmful content have been reported in literature and media \cite{Yu2022-ed, dalle}. However, failure modes regarding quality-of-service harms (i.e., not being able to generate an image for cultural concepts) have not been discussed to our knowledge.
\begin{table*}[]
\caption{Sample proof-of-concept FMEA, demo use}
\label{tab:fmea-use}
\centering
\resizebox{\textwidth}{!}{
\begin{tabular}{p{.12\textwidth}p{.11\textwidth}p{.25\textwidth}p{.24\textwidth}p{.24\textwidth}p{0.24\textwidth}}

\toprule
\textbf{Function} &
  \textbf{Type of failure mode} &
  \textbf{Failure mode} &
  \textbf{Effect} &
  \textbf{Cause} &
  \textbf{Control} \\ \hline
Artist \par selects image &
  Loss &
  Artist cannot select an appropriate image from the generated set of images &
  Dissatisfied  with generated images &
  Image quality is poor, the images do not match the prompt, the images are not   inspirational &
  User   testing/reporting \\ \cline{2-6}
 &
  Partial &
  Artist can only find a few appropriate pictures &
  Dissatisfied with generated images &
  Image quality is poor, the images do not match the prompt, the images are not   inspirational &
  User   testing/reporting \\ \cline{2-6}
 &
  Unintended &
  Artist selects images that could potentially be harmful for a given audience &
  Dissatisfied with generated images &
  Safety filters did not work as desired, challenges in the training process &
  User   testing/reporting \\ \bottomrule
\end{tabular}}
\end{table*}

\paragraph{\textbf{\textit{STPA}}}
Starting from the same set of losses identified for the data processing scope, we identified three potential hazards for our identified system in the use scope. The goal of our system is to support an artist in creating a video for a client using a T2I demo.

\begin{figure}[h]
  \centering
    \caption{The control structure diagram, demo use}
   \label{fig:control-use}
  \includegraphics[width=\linewidth]{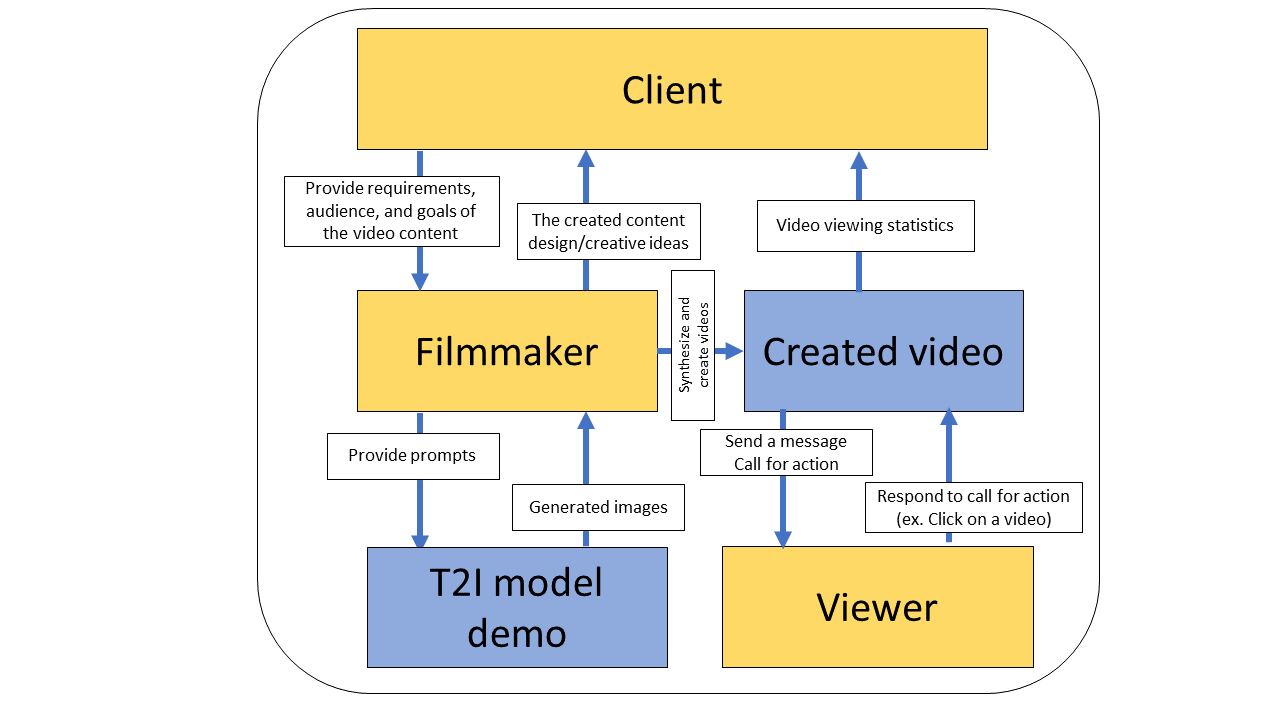}
\end{figure}

\begin{itemize}
    \item H1: The artist cannot create a video. (L1, L3, L6) 
    \item H2: The artist cannot create a video that meets client requirements. (L1, L3, L5, L6) 
    \item H3: The video disseminates false and harmful information. (L2, L3, L5, L6) 
\end{itemize}
A potential configuration of the control structure (illustrated in Figure \ref{fig:control-use} includes three human controllers (i.e., the client, filmmaker, and the viewer) and two automated controllers (i.e., the T2I model demo and the created video). For the purpose of this analysis, we focus on the 3 control actions between the 3 human controllers and the T2I model demo (denoted in boxes with down arrows in Figure \ref{fig:control-use}): (1) Provide requirements/goals/audience for the video content; (2) provide prompts; (3) send a message. We then identified a non-exhaustive list of 9 \acp{UCA}. Table \ref{tab:stpa-use} illustrates 2 of these \acp{UCA}. The UCAs included miscommunication regarding the video content's requirements, audience, and goals, incorrect prompt design, and inappropriate/misinformed calls to action for the public/viewer.  

Similar to prior scopes, many of the \acp{UCA} lead to multiple hazards. From the three hazards we are considering, H1 and H2 primarily present performance-related risks; H3 presents social and ethical risks. Some of the UCAs we identified have been discussed in recent publications (e.g., how certain text prompts do not generate well-aligned images using these T2I models \cite{Dallery_gallery2022-kt, Wang2023-ts} and how T2I demos could create misinformation \cite{disinfo2022}). However, few UCAs identified in this proof-of-concept analysis have not been directly discussed in the literature. For example, the \ac{UCA} about how these T2I models work well for artists from specific socioeconomic conditions is not well-discussed in the literature. Moreover, the \acp{UCA} about the roles/ expectations of the client in the artists' practice and use of T2I demos is not discussed in the literature to our knowledge.


\begin{table*}
\caption{Sample proof-of-concept STPA, demo use}
\label{tab:stpa-use}
\LARGE
\resizebox{\textwidth}{!}{

\begin{tabular}{lllllllll}
\toprule
\textbf{Controller} &
  \textbf{Controlled process} &
  \textbf{Control action} &
  \textbf{Type} &
  \textbf{Context} &
  \textbf{H1} &
  \textbf{H2} &
  \textbf{H3}  &
  \textbf{H4} \\ \hline
Client &
  Filmmaker &
  Provide   requirements, audience and goals &
  Too   late &
  Client provides the requirement later than expected. & True &
  True &
  False &
  True
  \\ \hline
Filmmaker &
  T2I demo &
  Provide   prompts &
  Not   providing &
  Film maker cannot provide prompts because their choice of prompts is blocked &
  True &
  True &
  True &
  False \\ \bottomrule
\end{tabular}}
\end{table*}

\section{Discussion}
\label{discussion}
Safety engineering frameworks were originally designed for safety-critical systems where a failure/hazard could lead to significant injury or damage to a person, property, or environment (i.e., nuclear power plants, medical devices, airplanes) \cite{Dekker2019-sk}. The types of harms (i.e., sociotechnical harms) \cite{Shelby2022-oi} and technical systems (i.e., complex ML models) in the current conversation of responsible ML development are different from those typically considered in safety engineering. However, the practices and processes of safety engineering could bring a much-desired mature and systematic perspective to responsible ML development \cite{Rismani2022-ao, Madaio2022-yz, Dobbe2022-kf}.  Through a case study of the development and use of T2I demos, we explored application of two safety engineering frameworks along the ML development pipeline and examined if we could discover failures and hazards that could lead to social and ethical risks. 

\subsection{Safety engineering perspective: The value of analyzing processes and interactions }
Our findings illustrate a potential approach for applying failure and hazard analysis tools from safety engineering to examine different scopes along the ML development pipeline. In total, we were able to identify 50+ potential hazards and failures across the three scopes of analysis without having any details or assumptions about the type of model used in the T2I demo. The identified failures and hazards covered many different issues and topics corresponding to the three stages of the ML development pipeline. Moreover, we were able to identify hazards and failures that could present \textit{social }and\textit{ ethical risks} without making any changes to the original safety engineering framework. This signals the potential usefulness of safety engineering for responsible ML development practices. 

Responsible AI assessments have often focused on assessing the behavior of a single model with respect to AI ethics principles, such as fairness and transparency \cite{Wang2022-ck, Tahaei2023-hp}. Recently, there has been movement towards understanding processes and interactions involved with the development and deployment of \ac{ML} systems in the Human-Computer Interaction and Science and Technology Studies communities where scholars have investigated harms emergent from human-AI interaction \cite{Wang2022-cx,Blodgett2022-db}. Similarly, Responsible AI scholars have examined data collection processes and pointed out clear areas for improvement \cite{sambasivan2021re, Hutchinson2020-jm,Denton2021-ol}. The hazards and failures found from applying FMEA and STPA to interactions and processes along the ML development pipeline, reiterate findings from this related work and supports the potential value of translating safety engineering practices for responsible \ac{ML} development. Moreover, compared to current responsible AI assessments such as algorithmic impact assessments and third party ML auditing, the aforementioned safety engineering frameworks support a proactive approach to systematically analyze a system's failures and hazards at a pragmatic level and early on in the development process.   

The FMEA process was especially comprehensive in analyzing \textit{processes}. The FMEA analysis facilitates the discovery of failure modes by observing and examining each of the steps in detail and focusing the analysts on being introspective about their current practices and seeing how they could be improved \cite{Carlson2012-yh,Rismani2021-qy,Raji2020-dw}. The STPA process pushes analysts to understand \textit{interactions} between different parts of a system, and how they could break down \cite{Leveson2018-no}. This can bring much value to analyzing ML systems as it is common for aspects of the ML pipeline to be siloed, in which people working on different components of the system have limited interaction and may not observe issues that could arise when a piece of information/data/design is passed on from one group to another. Bringing the safety engineering mindset to responsible AI practices would allow practitioners to look "beyond the model" and into processes and interactions. This shift in focus can help identify potential failures and hazards even when analysts do not fully understand the model’s capabilities. In this case, we found that applying the same frameworks at three different stages of the ML development pipeline created dialogue between siloed teams across the full product life cycle that would likely result in the deployment of a safer system. This facilitates a more systematic approach for responsible \ac{ML} by building on 100 years of designing safe systems.
\subsection{Fostering safety culture in responsible ML}
In safety engineering practice, dedicated individuals are responsible for conducting failure and hazard analysis \cite{Dekker2019-sk}. These experts could also be made responsible for technology development \cite{Carlson2012-yh,Dekker2019-sk}. Recognizing the organizational challenges of implementing responsible ML practices, the movement towards safety engineering practices and frameworks must be accompanied by shifts in organizational structure, incentives and towards increased internal capacity for conducting this type of analysis \cite{Rakova2021,Rismani2022-ao,Martelaro2022}. Integrating safety engineering frameworks into a company's workflow could take a long time, and it requires commitment/buy-in from leadership. It is important to establish when and who should be responsible for various roles such that appropriate incentive and compensation structures are put into place. 

FMEA-like processes could be done by system experts who have a good understanding of a specific product or process. For example, the group in charge of data processing for training data could conduct an FMEA analysis analogous to the one in our case study, and they would have the necessary information to conduct such an analysis. Whereas STPA analysis could be effectively performed by practitioners who have a system-level perspective of the varying range of key \ac{ML} models and stakeholders involved so that they can map out an appropriate control structure. Both FMEA and STPA-like analysis could be done earlier in the \ac{ML} development process as long as there is a basic understanding of the system/process design. It would be beneficial to start STPA processes earlier than FMEA since making changes to the system dynamics are often harder than modifying components/steps themselves. Both of these documents need to be updated as significant system changes are implemented or when there are new findings from a growing body of research on \ac{ML} systems and unexpected failures or near misses.  

\subsection{Limitations and opportunities for applying FMEA and STPA for ML products}
One of the key limitations of our analysis is that we have not considered all the possible scopes (i.e. T2I model architecture). Considering how generative T2I models are structured and set up it is challenging to break down the model architecture into specific components or model them as a control structure. This is an active line of research that we are investigating and invite the community to consider as a potential avenue for assessing \ac{ML} models. Secondly, the FMEA and STPA analysis depend heavily on how an analyst sets up the scope of analysis and maps the system/process. If the system is not mapped out adequately and there are false assumptions made, then the analysis will not be comprehensive or accurate. We validated our findings by comparing them to the current conversation in the literature. However, a comprehensive validation of this type of analysis is hard because we are trying to predict potential failures/UCAs that have not occurred and therefore, these types of analysis are meant to be iterative documents/processes. To show the validity of this type of analysis, we call on the FATE community to assess the use and applicability of such frameworks and critique the quality of these assessments so that we can ultimately build best practices around the application of such analysis frameworks. Another key limitation of this analysis is the analysts’ positionality (i.e., lived experiences), biases (i.e., professional expertise), and incentives (i.e., a company culture that promotes fast launches) which could impact quality, coverage, and the time they spent doing the analysis and quality and coverage. This type of analysis will only be meaningful in a company and technology regulatory ecosystem that emphasizes responsible \ac{ML} and due diligence practices.

\section{Conclusion}
Recognizing the rapid pace of movement towards incorporating generative T2I models in product development in creative practices, we examined the insights safety engineering frameworks, such as STPA and FMEA, could provide for responsible development and integration of such models within the creative practice. Our analysis underscored some of the existing concerns identified in the literature and highlighted potential novel areas of concern that could be further examined. Our case study highlights the value safety engineering analysis frameworks could provide for responsible \ac{ML} development and highlighted the value of looking beyond a single model and considering processes and interactions. 

\begin{acks}
We thank Freya Salway, who helped us organize the workshops, initiated the connections and invited the artists. We are grateful to our interview and workshop participants for taking the time to share their experiences, expertise, and feedback. We thank Remi Denton, Kathy Meier-Hellstern, Mohammad Havaei, and Tim Falzone for sharing their expertise with us.
We also thank our anonymous reviewers for their feedback on this paper. 
Finally, this work was financially supported by the Natural Sciences and Engineering Research Council of Canada and lead author's part-time internship at Google Research.
\end{acks}
\bibliographystyle{ACM-Reference-Format}
\bibliography{mainref}
\begin{appendices}

\newpage
\onecolumn
\section{Supplementary Material}
\label{sec:supplementary_methods}
\subsection{Interview protocol - T2I ML practitioners}

Thank you for taking the time to participate in this interview. 

I am conducting these expert interviews to understand how current T2I models are developed and evaluated. The data from this study will be used to examine potential social and ethical failures/hazards that could emerge from development of such models and their use as an assistive tool in the art creation process. 

I will start by asking you about your role in developing T2I models and then further delve into how you went about curating the data for training, creating the models and evaluating them. My goal is to have an in-depth understanding of the ML development pipeline for these models.  

Before we begin, please note that in responding to these questions, you should not share any confidential or proprietary information or any information that may be subject to any non-disclosure agreement, employment agreement, or otherwise. If you are unable to answer any of the questions without sharing confidential or proprietary information, please skip the question. Any information you provide in response to this interview will be deemed to be non-confidential and non-proprietary.

\begin{itemize}
    \item Could you briefly describe your role and current responsibilities? 
    \item What type of T2I model have you worked on in the past? 
    \item How were you involved in the development and evaluation (qualitative and quantitative) of the model? 
    \item When did you start working on this project? At what point did you enter the development process? (i.e. after training data was curated after the model architecture was selected)
    \item How did you contribute to the development  and evaluation of the T2I model?
    \item Could you walk me through your development and/or evaluation process when working on [based on answer to previous questions]?  
    \item How did you collect the data for training? 
    \item How did you select/filter the data for training? 
    \item Could you elaborate on how you went about training the model? 
    \item How did you decide on regularization? hyperparameters/ 
    \item When choosing a model architecture, what factors did you consider? 
    \item What is the variability of the model based on different initializations or subsets of data? Or different architectures (e.g. different text encoders)  
    \item What do you evaluate? How did you evaluate it?
    \item What is the metric that you selected? 
    \item Why did you select this method? 
    \item What are the limitations of this evaluation method/metric? 
    \item What is your evaluation of T2I for underrepresented features? 
    \item What is the importance/difficulty of different prompts? 
    \item Who did you work and collaborate with throughout this process? (ex. Crowd workers, product manager) 
    \item How did you work with them? 
    \item Tease at each one of the roles and contributions. 
    \item Who were the key decision makers in the development process? How did you communicate with them?
    \item What kind of decisions were they making? 
    \item How were you involved in those decision-making processes?

\end{itemize}

\subsection{Workshop with artist protocol}

\textbf{Introduction and welcome [15 min]} 

During today’s workshop we will be facilitating three main conversations. We will use the information and knowledge generated from these conversations to examine how we can improve the design and deployment of T2I models for use by digital artists. We will cover one main theme for each of the conversations. These are: 

\begin{itemize}
    \item Use of T2I models in art creation process
    \item Potential harms of T2I models in art creation 
    \item Looking forward: improving T2I models for art creation

\end{itemize}
Each conversation will start with a primer presentation and we will follow with a series of discussion questions. 

This will be an interactive workshop and we are excited to learn about your experience and expertise. 

\subsubsection{Conversation 1: Use of T2I in Art Creation [75 min]}
Speaker to provide a 10-min talk summarizing the discussions from the first day. 

List of questions to cover during this conversation: 
\newline
\textbf{
Art creation process (15 min)
}
\begin{itemize}
    \item Can you describe your work as an artist and the approach that you take to create art? 
    \item What type of art do you create? (what mediums do you use?)
    \item What types of digital tools do you use? (e.g., Illustrator, digital film, etc.)
    \item Who is your main audience as a creator? 
    \item Who do you work with in this process? 

\item What are the top 3 values that you exercise and care about as an artist? 
\item Write a brief 1-2 sentence definition of these values.
\item Can you talk about why these values are important?

\end{itemize}

\textbf{Use of T2I in the art creation process (20 min)}
\begin{itemize}
    \item  How do you currently use AI technology in your creation process? More specifically:
How about T2I models? 
Do you see T2I models as similar or different from existing digital tools? How so?
What do you like (is appealing to you) about T2I models? Why use them (what motivates you to use them)?
Can you draw your process of using T2I models in your art creation practice? 
Think about who is involved and when. 
How does the use of T2I model feed into your final art piece? 
Does this art creation process change at all when you aren’t using T2I models? What would you use instead of the T2I model?
Reflecting on your values, how does use of T2I models amplify, hinder, or complicate these values?

\end{itemize}

\textbf{Advantages, failures and limitations of T2I (20 min)}
\begin{itemize}
    \item  
When thinking about use of T2I models:
\item When have these systems not worked well/failed for you? 
\item What was the nature of the failure? (e.g., could not render what you wanted? Rendered something problematic?)
\item On a scale from 1-10, how adversely did this impact you or your creative process?
\item Did it make you consider *not* using T2I models?
\item Have you ever leveraged these failures as part of your artistic process? 
When have they worked well for you? 
\item On a scale from 1-10, how did this positively impact you or your creative process?
\item What are the advantages of using T2I models in your creation process?
\item What advantages are most important to you?
\item What are the disadvantages/limitations of using T2I models in your creation process? 
\item What limitations are most impactful for you?

\end{itemize}

\textbf{Debrief (5 min)
}
\begin{itemize}
    \item Now that we have talked about your use of T2I models we want to explore two main concepts: potential harms and opportunities of such models from your perspective.

\end{itemize}
    
\subsubsection{Conversation 2: Harms and harms reduction [90 min]} 

In this presentation we shared the most commonly discussed harm types when it comes to integration of algorithmic systems in different social systems. During our next conversation we want to explore if, how and when these harms or (any other ones) emerge when using T2I models within the art creation process. 

List of questions to cover during this conversation:

\textbf{Introductory questions (15 min):
}

\begin{itemize}
    \item What potential harms might arise in the art creation process when T2I models *are not* used? 
\item Who might be harmed in the art creation process?
\item Who might be harmed by the created art piece? 
\item What are your first impressions on the presented harm taxonomy? 
\item Did any of the presented harms connect to your experience of use of technology/algorithms in the art creation process?
\item Which ones were the most impactful?

\item When thinking through the use of technology in the art creation process, how do you think about potential harms and who is impacted? 
\item Who might be harmed in the art creation process? [Probe on communities, job role]
\item Who might be harmed by the created art piece?

\end{itemize}

\textbf{Thinking through potential harms (45 min):
}
\begin{itemize}
    \item Take 10 minutes to think through each of the five categories of sociotechnical harm and prompt a brainstorming exercise on each one. Divide into 5 groups and each group focuses on one category of harms.  

\item Thinking about your experience of using T2I models in the art creation process what harms do you specifically see emerging when it comes to:  write 3-5 sentences describing the scenario.
\item What do you think is the primary source of these harms (technologies, structures, policies, processes, practices) - intentionally or unintentionally?
\item What type of harms are missing from the presented taxonomy based on your experience?

\end{itemize}
\textbf{Harms reduction strategies (15 min): 
}
\begin{itemize}
    \item Which 1-2 harms do you think are most important to minimize or eliminate? Why? 
    \item What are some potential ways of mitigating the discussed harms? 
    \item What communities need to be involved to help mitigate these harms ?

\end{itemize}

\subsubsection{Conversation 3: Looking Forward [60 min]} 

Now that we have discussed potential harms and harm reduction strategies that could emerge from use of T2I models in the art creation process for different stakeholders. We will turn our attention to the future use of these models.

\textbf{Exploring opportunities:}
\begin{itemize}
    \item Are there social and cultural problems that T2I models could help alleviate? 
    \item Could we help in better representation of underrepresented cultures in art using these tools?
    \item How can we use these tools for artists' social benefits? How could artists who are not using these tools get benefit?
 
\end{itemize}

\textbf{Building futures}

\begin{itemize}
    \item Blue sky wants/needs (sets context for later prompts, and allows participants to identify values outside of existing structures; allows peers to engage with each other and open process of potential value conflicts)
\item Lay out a vision of what an ideal T2I model could be
I want to see the data it was trained on
I want the probe to work better
I want XYZ
\item Engage with at least two other participants’ visions

\end{itemize}
Prioritization and role-taking (final reflection, focusing on how emergent values from prior conversations can be incorporated into T2I design).
If we were to launch the T2I we’ve been talking about tomorrow, what would it look like, how would it integrate into your creative practice, and what role do you see yourself playing on this new platform?
What else needs to be in place

\textbf{
Closure activity: give an overview of the whole session. 
}
\newpage
\subsection{Information gathering flow diagram}

\begin{figure}[ht]
  \centering
    \Description{Sources of information for the case study} 
      \caption{Sources of information for the case study}
  \label{fig:info}
  \includegraphics[width=\linewidth]{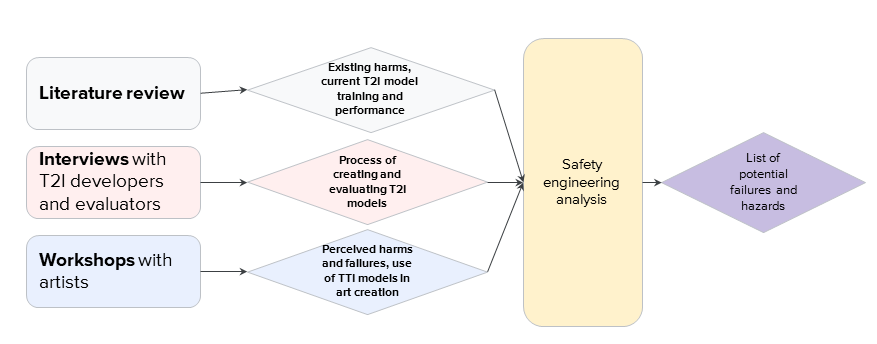}
\end{figure}

\subsection{FMEA and STPA tables}
\begin{table} [ht]
\caption{Sample proof-of-concept FMEA, data processing}
\label{tab:fmea-data}
\LARGE
\resizebox{\columnwidth}{!}{

\begin{tabular}{p{.15\textwidth}p{.1\textwidth}p{.15\textwidth}p{.2\textwidth}p{.05\textwidth}p{.15\textwidth}p{0.01\textwidth}p{0.2\textwidth}p{.01\textwidth}p{.01\textwidth}}
\toprule

\textbf{Function} &
  \textbf{Type of failure mode} &
  \textbf{Failure mode} &
  \textbf{Effect} &
  \multicolumn{1}{l}{\textbf{Severity}} &
  \textbf{Cause} &
  \multicolumn{1}{l}{\textbf{Occurrence}} &
  \textbf{Control} &
  \multicolumn{1}{l}{\textbf{Detection}} &
  \multicolumn{1}{l}{\textbf{RPN}} \\ \hline
\textbf{Filter records identified as containing sensitive data} &
  Loss &
  Sensitive data is not filtered and included in the training set. &
  Sensitive data might be reproduced by the model output. &
  10 &
  There was no requirement to filter the data for sensitive data. &
  4 &
  Model development team checks that the data was filtered for sensitive information prior to starting the training. &
  2 &
  80 \\ \hline
\textbf{} &
  Partial &
  Sensitive data is not adequately filtered and it is included in the training set. &
  Sensitive data is not filtered and it is included in the training set. Sensitive data might be reproduced by the model output. &
  9 &
  Wrong filter thresholds are set. &
  7 &
  Model development team monitors the outputs of the models for sensitive data generation. &
  5 &
  315 \\ \hline
\textbf{} &
  Exceeding &
  Non-sensitive data is also filtered. &
  The resulting dataset for training is too small. &
  8 &
  Wrong filter thresholds are set. &
  7 &
  Model development teams monitors how much data is filtered out after this one filter. &
  2 &
  112 \\ \bottomrule
\end{tabular}}
\end{table}

\begin{table}[ht]
\caption{Sample proof-of-concept FMEA, model integration}
\label{tab:fmea-model}
\LARGE
\centering
\setlength{\tabcolsep}{4.5pt}
\resizebox{\linewidth}{!}{
\begin{tabular}{p{.1\textwidth}p{.1\textwidth}p{.15\textwidth}p{.25\textwidth}p{.05\textwidth}p{.25\textwidth}p{0.01\textwidth}p{0.25\textwidth}p{.01\textwidth}p{.01\textwidth}}

\toprule
Function &
  Type   of failure mode &
  Failure   mode &
  Effect &
  \multicolumn{1}{l}{Severity} &
  Cause &
  \multicolumn{1}{l}{Occurrence} &
  Control &
  \multicolumn{1}{l}{Detection} &
  \multicolumn{1}{l}{RPN} \\ \midrule
Filter   text prompts &
  Degradation &
  Filter does not work for novel words &
  Model does not filter prompts containing inappropriate language. &
  6 &
  Safety filters are not updated/improved over time. &
  7 &
  Product team monitors user feedback to identify potential negative feedback on prompt filters and complaints about the model accepting new and inappropriate prompts. &
  7 &
  294 \\ \hline
 &
  Partial &
  Filter works for English words written in Latin letters (e.g., it does not work when the Arabic language is written using Latin words) &
  Users enter harmful prompts using non-English phrases written with Latin letters &
  6 &
  Filter allows text prompts containing non-English words &
  7 &
  Product team monitors user feedback to identify potential negative feedback on how their prompts were filtered and complaints about the model accepting new and inappropriate prompts. &
  7 &
  294 \\ \hline
 &
  Exceeding &
  Seemingly   appropriate prompts are rejected for no good reason. &
  Users   cannot enter what they would like to enter &
  7 &
  Filters automatically eliminate word combinations that are not directly harmful but are in some form correlated to content that is marked as   harmful. &
  5 &
  Product team monitors user feedback &
  7 &
  245 \\ \bottomrule
\end{tabular}}
\end{table}

\begin{table}[ht]
\caption{Full list of UCAs from proof-of-concept STPA, data processing integration}
\label{tab:full-stpa-data}
\centering
\resizebox{\columnwidth}{!}{
\begin{tabular}{p{.12\textwidth}p{.11\textwidth}p{.21\textwidth}p{.1\textwidth}p{.24\textwidth}p{.05\textwidth}p{0.05\textwidth}p{0.05\textwidth}}
\textbf{Controller} &
  \textbf{Controlled Process} &
  \textbf{Control Action} &
  \textbf{Type} &
  \textbf{Context} &
  \textbf{H1} &
  \textbf{H2} &
  \textbf{H3} \\ \hline
Reponsible ML team &
  Input filter &
  Set threshold for pre-designed filter for the prompt filter &
  Not providing &
  The filter thresholds are not updated by the RML for this product. Old filter thresholds are used. &
  FALSE &
  TRUE &
  TRUE \\ \hline
Reponsible ML team &
  Input filter &
  Set threshold for pre-designed filter for the prompt filter &
  Providing &
  Inappropriate filter thresholds are provided by RML &
  FALSE &
  TRUE &
  TRUE \\ \hline
Reponsible ML team &
  Input filter &
  Design new filters for the prompt filter &
  Too late &
  The new filters are not designed in time for product deployment &
  FALSE &
  FALSE &
  TRUE \\ \hline
Reponsible ML team &
  Input filter &
  Design new filters for the prompt filter &
  Providing &
  The new filters are filtering out most of the prompts that users would want to put in &
  FALSE &
  TRUE &
  FALSE \\ \hline
Product team &
  Model &
  Set model design parameters during fine-tuning/re-training of the model for the specific application &
  Not providing &
  The model parameters lead to poor quality images and lack of text to prompt matching. &
  TRUE &
  FALSE &
  FALSE \\ \hline
Product team &
  Model &
  Set model contraints for what it can generate in terms of the type of shapes, colors, etc &
  Providing &
  The model constraints does not allow the user to create types of images that they desire &
  FALSE &
  TRUE &
  FALSE \\ \hline
Reponsible ML team &
  Output fitler &
  Set threshold for pre-designed filter for the image filter &
  Not providing &
  The filter thresholds are not updated by the RML for this product. Old filter thresholds are used. &
  FALSE &
  FALSE &
  TRUE \\ \hline
Reponsible ML team &
  Output fitler &
  Design new filters for the image filter &
  Too late &
  The RML provides the new filters too late for the product team. &
  FALSE &
  FALSE &
  TRUE \\ \bottomrule
\end{tabular}}
\end{table}

\begin{table}[ht]
\caption{Full list of UCAs from proof-of-concept STPA, model integration}
\label{tab:full-stpa-model}
\centering
\resizebox{\columnwidth}{!}{
\begin{tabular}{p{.12\textwidth}p{.11\textwidth}p{.21\textwidth}p{.1\textwidth}p{.24\textwidth}p{.05\textwidth}p{0.05\textwidth}p{0.05\textwidth}}
\textbf{Controller} &
  \textbf{Controlled Process} &
  \textbf{Control Action} &
  \textbf{Type} &
  \textbf{Context} &
  \textbf{H1} &
  \textbf{H2} &
  \textbf{H3} \\ \hline
Reponsible ML team &
  Input filter &
  Set threshold for pre-designed filter for the prompt filter &
  Not providing &
  The filter thresholds are not updated by the RML for this product. Old filter thresholds are used. &
  FALSE &
  TRUE &
  TRUE \\ \hline
Reponsible ML team &
  Input filter &
  Set threshold for pre-designed filter for the prompt filter &
  Providing &
  Inappropriate filter thresholds are provided by RML &
  FALSE &
  TRUE &
  TRUE \\ \hline
Reponsible ML team &
  Input filter &
  Design new filters for the prompt filter &
  Too late &
  The new filters are not designed in time for product deployment &
  FALSE &
  FALSE &
  TRUE \\ \hline
Reponsible ML team &
  Input filter &
  Design new filters for the prompt filter &
  Providing &
  The new filters are filtering out most of the prompts that users would want to put in &
  FALSE &
  TRUE &
  FALSE \\ \hline
Product team &
  Model &
  Set model design parameters during fine-tuning/re-training of the model for the specific application &
  Not providing &
  The model parameters lead to poor quality images and lack of text to prompt matching. &
  TRUE &
  FALSE &
  FALSE \\ \hline
Product team &
  Model &
  Set model contraints for what it can generate in terms of the type of shapes, colors, etc &
  Providing &
  The model constraints does not allow the user to create types of images that they desire &
  FALSE &
  TRUE &
  FALSE \\ \hline
Reponsible ML team &
  Output fitler &
  Set threshold for pre-designed filter for the image filter &
  Not providing &
  The filter thresholds are not updated by the RML for this product. Old filter thresholds are used. &
  FALSE &
  FALSE &
  TRUE \\ \hline
Reponsible ML team &
  Output fitler &
  Design new filters for the image filter &
  Too late &
  The RML provides the new filters too late for the product team. &
  FALSE &
  FALSE &
  TRUE \\ \bottomrule
\end{tabular} }
\end{table}

\begin{table}[ht]
\caption{Full list of UCAs from proof-of-concept STPA, model integration}
\label{tab:full-stpa-model}
\centering
\resizebox{\columnwidth}{!}{
\begin{tabular}{p{.12\textwidth}p{.11\textwidth}p{.21\textwidth}p{.1\textwidth}p{.24\textwidth}p{.05\textwidth}p{0.05\textwidth}p{0.05\textwidth}p{0.05\textwidth}}
\textbf{Controller} &
  \textbf{Controlled Process} &
  \textbf{Control Action} &
  \textbf{Type} &
  \textbf{Context} &
  \textbf{H1} &
  \textbf{H2} &
  \textbf{H3} &
  \textbf{H4} \\ \hline
Client &
  Filmmaker &
  Provide requirements, audience and goals &
  Not providing &
  The client does not communicate clear requirements for the artist &
  TRUE &
  TRUE &
  FALSE &
  TRUE \\ \hline
Client &
  Filmmaker &
  Provide requirements, audience and goals &
  Providing &
  The client provides overly restrictive requirements &
  TRUE &
  TRUE &
  TRUE &
  TRUE \\ \hline
Client &
  Filmmaker &
  Provide requirements, audience and goals &
  Too late &
  The client provides the requirement later than expected. &
  TRUE &
  TRUE &
  FALSE &
  TRUE \\ \hline
Filmmaker &
  T2I model UI &
  Provide prompts &
  Not providing &
  The film maker cannot provide prompts because their choice of prompts is blocked &
  TRUE &
  TRUE &
  TRUE &
  FALSE \\ \hline
Filmmaker &
  T2I model UI &
  Provide prompts &
  Providing &
  The filmmaker provides prompts that do not create very interesting images &
  TRUE &
  TRUE &
  TRUE &
  FALSE \\ \hline
Filmmaker &
  T2I model UI &
  Provide prompts &
  Out of order &
  The filmmaker provides prompts that do not have the right order of words for the given T2I UI &
  TRUE &
  TRUE &
  TRUE &
  FALSE \\ \hline
Filmmaker &
  T2I model UI &
  Provide prompts &
  Stopped too soon &
  The artist stops entering more prompts after the first few prompts did not result in desired images &
  TRUE &
  TRUE &
  TRUE &
  FALSE \\ \hline
Created video &
  Public/viewer &
  Send a call to action &
  Not providing &
  The call to action is not clear for the viewer &
  FALSE &
  TRUE &
  FALSE &
  FALSE \\ \hline
Created video &
  Public/viewer &
  Send a call to action &
  Providing &
  The call to action could cause unsafe behavior and spreads misinformation &
  FALSE &
  FALSE &
  FALSE &
  TRUE \\ \bottomrule
\end{tabular}}
\end{table}
\end{appendices}

\end{document}